\documentclass[runningheads]{llncs}
\usepackage{graphicx}
\usepackage{booktabs}
\usepackage{array}
\usepackage{xcolor}
\usepackage{colortbl}
\usepackage{tabularx}
\usepackage{multirow}
\usepackage{amsmath}
\usepackage{amssymb}
\usepackage{comment}
\usepackage{svg}
\usepackage[nointegrals]{wasysym} 
\usepackage{microtype}
\usepackage[hidelinks]{hyperref}
\usepackage{xurl}
\usepackage[most]{tcolorbox}

\newcommand{\rev}[1]{{\color{black}#1}}
\newcommand{\harsh}[1]{{\color{black}#1}}

\begin{document}

\title{Can Autonomous Vehicles Keep Their Cool? A Survey of Thermal Side-Channel and Perception Attacks}

\titlerunning{A Survey of Thermal Side-Channel and Perception Attacks}

%

\author{Harsh Aaryan\inst{1} \and
Ce Zhou\inst{1}\thanks{Corresponding author.} }

\authorrunning{H. Aaryan and C. Zhou}

\institute{
Missouri University of Science and Technology,
Rolla, MO 65409, USA\\
\email{\{hah3m,cezhou\}@mst.edu}
}


%

%

\maketitle              
\begin{abstract}
      Autonomous vehicles (AVs) couple safety-critical perception to thermally constrained compute. Physical adversaries can spoof thermal infrared (IR) sensors used for night driving, while unprivileged software co-tenants can fingerprint perception workloads from thermal, dynamic voltage and frequency scaling (DVFS), and thermal event interrupt (TEI) telemetry required for package safety. These paths differ in mechanism and privilege, but both exploit interfaces that AV stacks cannot disable without compromising Safety of the Intended Functionality (SOTIF). Existing microarchitectural and CPS surveys cover cache timing, adversarial ML, and camera or LiDAR spoofing; none frames physical thermal IR attacks and unprivileged silicon thermal telemetry as one residual-risk class. This survey systematizes thirteen works into thermal perception integrity, telemetry leakage, and model-information extraction, comparing them by privilege, timescale, co-tenancy, deployability, and AV relevance while mapping defenses to residual risk. A commodity proxy supports the software co-tenant branch only; physical IR evidence comes from cited literature. We identify open problems in on-target automotive evaluation, fusion trust under IR spoofing, chiplet thermal coupling, capability-based telemetry, and standards-ready threat analysis, motivating co-design across sensors, operating systems, silicon, and automotive standards.


\end{abstract}
\newcommand{\halfcirc}{\LEFTcircle}

\newtcolorbox{keyobs}[1]{%
  enhanced,
  colback=black!4,
  colframe=black!55,
  boxrule=0.55pt,
  arc=1.5pt,
  left=5pt,
  right=5pt,
  top=4pt,
  bottom=3pt,
  fontupper=\small,
  attach boxed title to top center={yshift=-2.2mm},
  boxed title style={
    colback=black!72,
    colframe=black!72,
    arc=7pt,
    boxrule=0pt,
    left=5pt,
    right=5pt,
    top=1.5pt,
    bottom=1.5pt
  },
  fonttitle=\bfseries\footnotesize\color{white},
  title={Key Observation #1},
  before skip=3pt,
  after skip=4pt
}

\section{Introduction}
Perception, planning, and control in autonomous vehicles (AVs) depend on multi-modal sensing with visible cameras, LiDAR, radar, and thermal infrared (IR) imagers, together with deep neural networks (DNNs) executed on high-performance GPUs and SoC accelerators~\cite{nvidia_drive,qayyum2020av}. Heat is an unavoidable byproduct of that computation. Platforms expose it through on-die sensors, dynamic voltage and frequency scaling (DVFS), and thermal event interrupts (TEIs), often readable without elevated privilege because thermal management must function under ordinary access controls~\cite{taneja2023,kim2022,zhang2024}. Separately, thermal IR cameras instrument the external heat field for night and low-visibility driving, where RGB contrast collapses~\cite{bhupathiraju2026}. In both cases, thermal interfaces were engineered for reliability or perception, not as confidentiality or integrity boundaries. Night perception cannot discard thermal IR without degrading the operational design domain, and package protection cannot disable thermal governors without risking thermal damage under Safety of the Intended Functionality (SOTIF).


Several trends compound this exposure. AV compute density and chiplet packaging intensify on-package thermal coupling; mixed-criticality stacks place infotainment and over-the-air (OTA) clients beside perception on shared silicon~\cite{autosar_adaptive,ros2}; and night driving elevates reliance on thermal IR. An adversary who never reads AV memory may still spoof pedestrians in the IR scene~\cite{zhu2021,wei2023,bhupathiraju2026}, or fingerprint which perception workload is running from operating-system telemetry~\cite{taneja2023,kim2022,gao2024}. \rev{These access paths differ in privilege and mechanism but expose complementary failures: untrusted heat patterns affect perception integrity, while readable telemetry affects workload confidentiality. Studying them together identifies residual risk across both boundaries and the shared constraint that defenses must preserve sensing and thermal control.}

Classical microarchitectural surveys cover cache timing, speculative execution, and power analysis in depth~\cite{ge2018survey,lyu2018survey,xiong2021survey,lou2021survey,spreitzer2018survey}, yet they rarely treat continuously readable thermal governors or thermal-camera pipelines as first-class AV security topics. Cyber-Physical Systems (CPS) and AV security surveys emphasize adversarial machine learning (AML) and sensor spoofing on cameras and LiDAR~\cite{humayed2017cps,qayyum2020av,eykholt2018robust,cao2019adversarial,shen2021fusion}, but thermal cameras appear mainly as an additional night-driving modality rather than as an attackable pipeline or as a source of OS-exportable workload telemetry. Energy-management security work such as CLKSCREW and VoltJockey addresses DVFS as a fault-injection surface~\cite{tang2017clkscrew,qiu2019voltjockey}; classic thermal covert channels establish heat as a communication medium~\cite{masti2015thermal,guri2015bitwhisper}. No prior survey jointly frames physical thermal IR spoofing and unprivileged silicon thermal/DVFS/TEI telemetry as one AV residual-risk class under a shared thermal safety envelope. This survey does not replace those literatures; it fills that gap by systematizing both access paths together, comparing them under one threat model, and showing why defenses that close only one path leave the other intact.

\harsh{\rev{This paper surveys thirteen representative works selected to compare thermal attack mechanisms, access requirements, and mitigation constraints relevant to autonomous vehicles.} We organize the works into physical thermal IR attacks, unprivileged silicon telemetry leakage, and energy-based model extraction. Attack family describes the mechanism and security objective, while adversary class describes access. We distinguish physical adversaries with proximity or line-of-sight to sensors or compute modules from unprivileged software co-tenants, such as infotainment or over-the-air (OTA) clients on shared compute. We compare the works by privilege, timescale, co-tenancy, deployability, and AV relevance, then examine defenses and remaining risks. Section~\ref{sec:evaluation} supplements the survey with an unprivileged telemetry evaluation on an Intel Xeon workstation with a Blackwell-class GPU. We use \emph{AV-analogue} for commodity DNN and CPU workloads that approximate perception compute intensity. \rev{These measurements test residual software-path leakage, channel substitution, and contention sensitivity}; they do not validate production AV stacks, automotive SoCs, or in-vehicle privilege policies. We conclude with design implications and open research problems.}

Our contributions are as follows:
\begin{itemize}
 \item \textbf{Taxonomy and Structured Comparison of Thermal Attack Paths.}
  We compare thermal IR spoofing, unprivileged thermal/DVFS/TEI telemetry, and energy-based DNN extraction along privilege, timescale, co-tenancy, deployability, and relevance-confidence axes, organized by physical and software co-tenant adversary classes.
  Unlike prior surveys that treat camera spoofing and microarchitectural leakage separately, this taxonomy makes residual risk visible across both access paths under one AV thermal safety envelope, enabling designers to identify which comparisons, metrics, and defenses transfer.

  \item \textbf{Defense Residual-Risk Synthesis Across Both Access Paths.}
  We map sensor, OS-telemetry, DVFS, packaging, and model controls to residual risk when thermal management must remain functional.
  For each control, the synthesis reveals what remains open on the other path and which mitigations conflict with SOTIF thermal constraints, a trade-off that single-path surveys do not surface.

  \item \textbf{Software-Path Proxy Characterization on Commodity Hardware.}
  \harsh{We measure software-path telemetry accessibility, workload fingerprinting, channel redundancy, observer sensitivity, and co-tenant stress on commodity hardware under an unprivileged policy.}
  It does not validate the physical IR branch or production AV stacks; it supplies a mechanism baseline for the software co-tenant class and shows that residual leakage is a telemetry class rather than a single locked interface.
\end{itemize}

\section{Background}
Modern AVs implement a layered sense--plan--act loop on thermally constrained embedded compute. The sensing layer typically fuses visible cameras, LiDAR, radar, ultrasonic sensors, and thermal IR imagers. Thermal IR is especially important in low-visibility conditions (night, fog, rain, glare), where pedestrians and cyclists may be detected by heat signature rather than by reflected visible light~\cite{bhupathiraju2026}. Raw frames pass through camera-specific pipelines for equalization, calibration, non-uniformity correction, and sometimes optical lensing compensation before entering DNN detectors and multi-modal fusion modules.

The perception outputs feed planning and control stacks that run alongside lower-criticality services (infotainment, logging, OTA update clients) on shared GPU- or SoC-class accelerators~\cite{nvidia_drive}. Automotive middleware such as AUTOSAR Adaptive~\cite{autosar_adaptive}, ROS~2~\cite{ros2}, and vendor SDKs (for example, NVIDIA DriveWorks~\cite{nvidia_drive}) often schedule mixed-criticality partitions on a common Linux-derived OS image. Safety-critical inference and non-safety processes therefore contend for the same silicon, caches, memory bandwidth, and thermal budget.

\begin{figure}[t!]
  \centering
  \includegraphics[width=\linewidth,trim={72 166 73 166},clip]{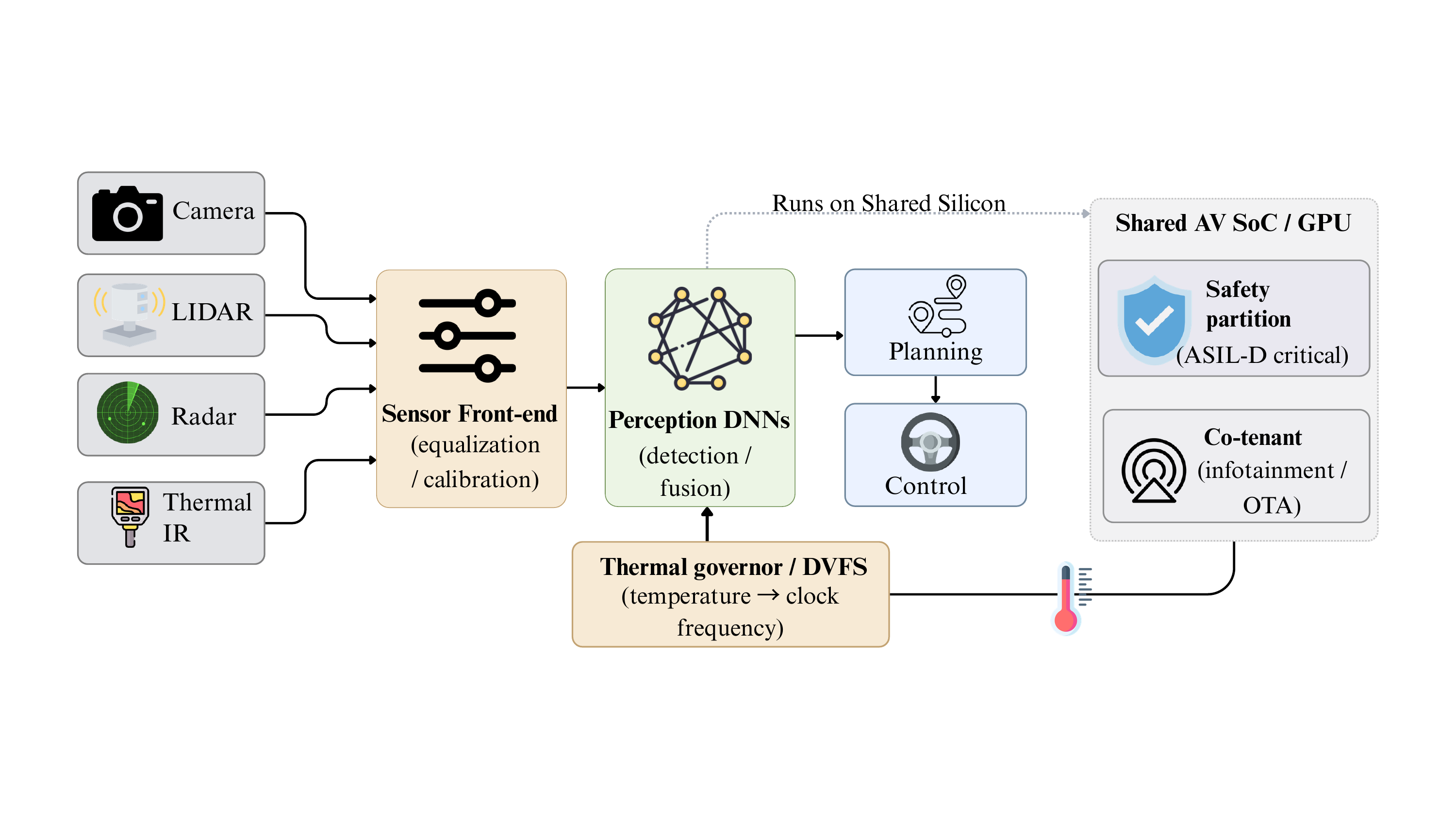}
  \caption{AV sense--plan--act stack on thermally constrained shared compute.
  Perception DNNs and lower-trust co-tenants (infotainment, OTA) share CPUs/GPUs under a common thermal governor.}
  \label{fig:av-stack}
\end{figure}

Figure~\ref{fig:av-stack} summarizes this pipeline: external sensing, perception DNNs, planning/control, and co-tenant software share one thermally managed compute substrate. Sustained inference pushes packages toward their thermal design point. When this occurs, DVFS and thermal governors reduce clocks to protect the SoC. The resulting temperature, power, and frequency signals are continuous control inputs for safe operation, not incidental diagnostics.

\subsection{Thermal Management and DVFS}
\label{sec:thermal-dvfs}

Thermal management closes a feedback loop among power dissipation, SoC temperature, and clock frequency. On-SoC digital thermal sensors and package sensors feed OS and firmware governors. When temperature or power limits are approached, the platform lowers voltage and frequency, throttles accelerators, or raises TEIs at thermal thresholds~\cite{zhang2024}. As perception latency and frame rate depend on sustained clocks, AV platforms cannot simply disable these mechanisms: doing so would risk thermal damage or violate intended functionality under ISO~21448 (SOTIF) thermal envelopes~\cite{iso21448}.

This loop creates an inherent observation channel. Workload and often data-dependent switching activity modulate power; power modulates heat; governors couple heat back into frequency and interrupt timing. Classic microarchitectural side channels (cache timing, speculative execution) exploit instruction pipeline artifacts~\cite{osvik2006cache,yarom2014flush,kocher2019spectre}. Thermal and DVFS channels instead exploit the energy-management substrate that every high-performance AV accelerator must expose. Energy-oriented attacks such as CLKSCREW~\cite{tang2017clkscrew} and VoltJockey~\cite{qiu2019voltjockey} treat DVFS as a fault-injection surface; this survey treats the same interfaces primarily as \emph{read} channels.

\subsection{Software-Readable Telemetry}

On commodity and embedded Linux platforms, temperature, frequency, and GPU metrics are commonly readable without elevated privilege through \texttt{/sys\allowbreak/class/thermal}, \texttt{hwmon}, CPU frequency sysfs, and vendor tools such as \texttt{nvidia-\allowbreak smi}~\cite{taneja2023,kim2022}. Intel Running Average Power Limit (RAPL) \texttt{energy\_uj} may be present but is often root-restricted; frequency and GPU power frequently remain open even when RAPL is locked~\cite{gao2024,wang2022}. Sampling rates range from sub-hertz thermal updates to tens of hertz for GPU queries, which is sufficient for workload fingerprinting and, in prior work, for architecture inference.

From an AV security standpoint, telemetry minimization conflicts with thermal safety: Automotive Safety Integrity Level (ASIL)-oriented partitions still need accurate thermal state. Partitioning therefore often hardens memory and CPU isolation while leaving thermal sysfs globally readable. That asymmetry is the operational premise for co-tenant software attacks surveyed.
\section{Threat Model}
\label{sec:threat-model}

\begin{figure}[t!]
  \centering
  \includegraphics[width=\linewidth,trim={107 70 101 69},clip]{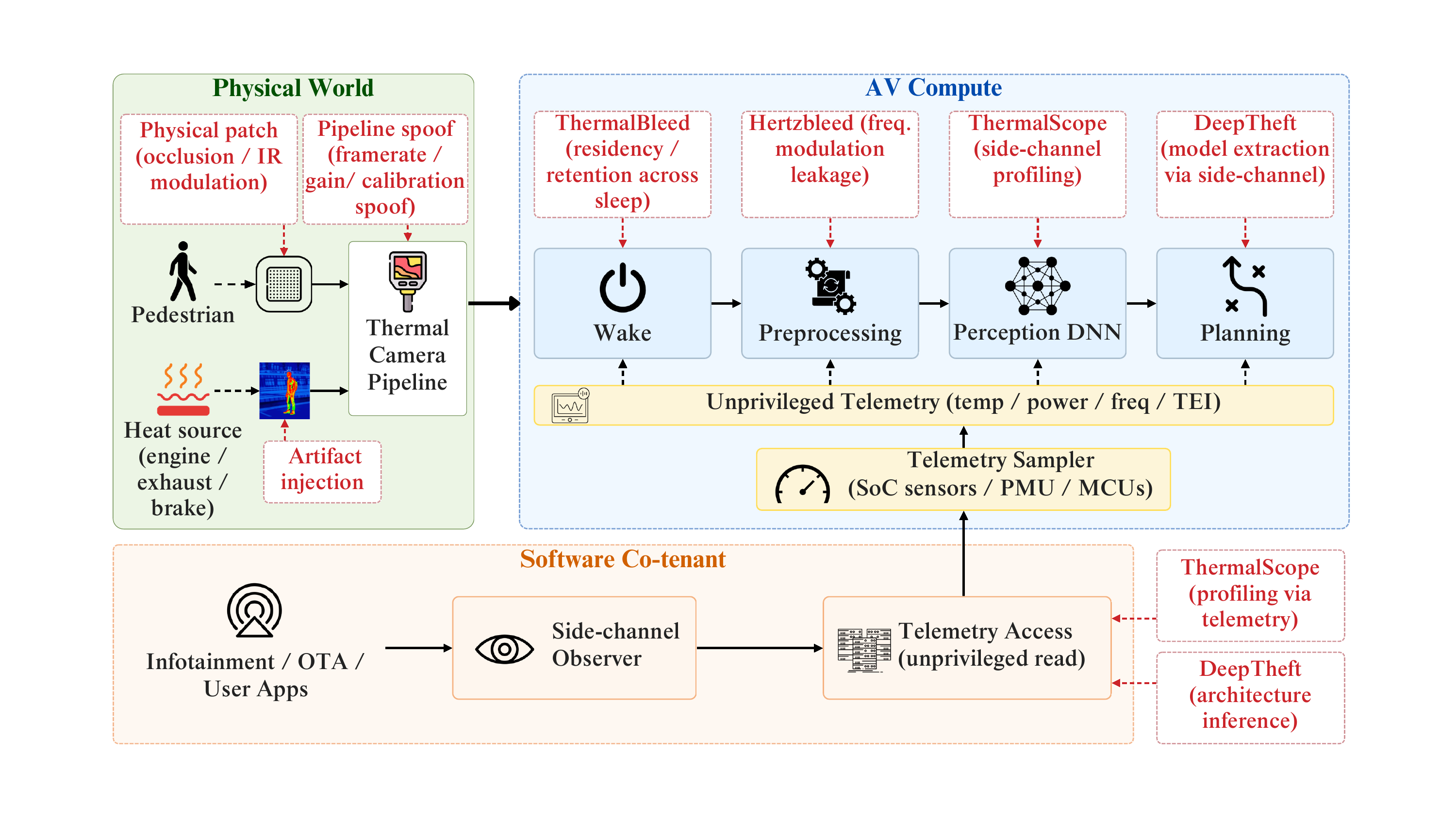}
  \caption{Threat model for thermal side-channel attacks on AVs.
  Physical adversaries manipulate the IR scene or camera pipeline; software co-tenants sample unprivileged temperature, power, frequency, or TEI timing on shared compute.}
  \label{fig:threat-overview}
\end{figure}

Figure~\ref{fig:threat-overview} summarizes the two-class threat model used throughout the survey. The classes differ in access path and privilege but exploit interfaces that AV platforms must keep operational. Thermal IR supports night perception, whereas thermal governors ensure package safety. Both classes can influence the sense--plan--act loop without requiring access to AV memory.

The \emph{physical adversary} has line-of-sight or short-range proximity and can inject or manipulate heat in the external environment, for example through a thermal decoy, bulb board, wearable hot/cold pattern, or a projector aimed at a thermal sensor~\cite{zhu2021,wei2023,bhupathiraju2026}. This attacker requires no access to the vehicle's compute platform and operates black-box against the perception pipeline. Success is measured by false detections, missed pedestrians, or planner-level reactions under night or low-visibility conditions where the stack trusts thermal evidence.

\harsh{Attack family denotes mechanism and objective; adversary class denotes location and access. They are orthogonal: physical or software adversaries may pursue different objectives through distinct thermal channels.}
The \emph{software co-tenant} is co-located on shared compute, either as a lower-privilege process (infotainment or OTA client) or as a user-level process that reads OS thermal and power telemetry (Linux \texttt{hwmon}, \texttt{/sys/class/thermal}, CPU frequency sysfs, or GPU APIs such as \texttt{nvidia-smi})~\cite{taneja2023,kim2022}. This attacker requires no root privilege and no performance-monitoring units. Success is measured by workload fingerprinting, covert signaling, or architecture/hyperparameter recovery from residual channels that remain after RAPL lockdown~\cite{gao2024,murdock2021}.

\noindent\textbf{Adversary Goals.}
In both classes, the adversary seeks to disrupt or exfiltrate safety-critical computation. On the physical path, the primary goals are perception integrity failures: false pedestrians or obstacles, suppressed true detections, and fused decisions that authorize unsafe maneuvers when RGB contrast is low~\cite{bhupathiraju2026,zhu2021,wei2023}. On the software path, the primary goals are confidentiality and reconnaissance: identifying which perception workload or model is running, recovering architecture or hyperparameter candidates, or establishing a low-rate covert channel from an unprivileged co-tenant~\cite{taneja2023,kim2022,zhang2024,gao2024}. Goals that require kernel compromise, physical EM probing, or TEI programming are treated as escalations beyond the default co-tenant model and are noted only when a cited work specifically assumes them~\cite{horvath2025}.

\noindent\textbf{Assumptions.}
We assume the software adversary can (i)~schedule an unprivileged process on the same package or SoC as a victim workload, (ii)~sample documented thermal, power, and frequency interfaces at OS-supported rates, and (iii)~train offline classifiers on labeled traces of known workloads. We assume the physical adversary can place or wear heat sources within sensor range, but not alter camera firmware or the detector weights. We do not assume kernel compromise, RAPL \texttt{energy\_uj} access, physical EM probes, or TEI-programming privileges unless a cited work specifically requires them. Automotive middleware (AUTOSAR Adaptive~\cite{autosar_adaptive}, ROS~2~\cite{ros2}, NVIDIA DriveWorks~\cite{nvidia_drive}) often rests on embedded Linux privilege models in which thermal sysfs and GPU telemetry remain readable to non-safety partitions. That exposure is typically retained for bring-up and field diagnostics, for default kernel hwmon and thermal-class permissions that treat temperature as a shared platform service, and for ASIL governors that must observe package state. OEM hardening of those interfaces is deployment-specific.

\section{Surveyed Attacks}
Access paths differ, yet the surveyed attacks exploit the same underlying fact: heat was instrumented for night perception or package reliability, not confidentiality. We first summarize the attack landscape and comparison axes, then cover physical manipulation of the IR scene and camera pipeline, unprivileged reads of on-package temperature, power, frequency, and TEI timing, and finally energy-side architecture and weight recovery. Defenses and metrics do not transfer across subclasses; the shared anti-pattern is a residual-risk frame for AV designers.

\subsection{Attack Landscape}
\label{sec:attack-landscape}
Table~\ref{tab:taxonomy} summarizes the surveyed works, and Table~\ref{tab:compare-axes} compares the same works by timescale, co-tenancy, qualitative deployability, and relevance confidence. 
\rev{The thirteen works were selected to compare physical thermal IR attacks, thermal/power/frequency/TEI leakage, and energy-based DNN extraction. They span scene and pipeline manipulation, continuous and thresholded observations, and workload identification through model recovery, supporting comparison of access requirements and mitigation constraints.} Classic cache and speculative-execution attacks~\cite{osvik2006cache,yarom2014flush,kocher2019spectre,lipp2018meltdown} are out of scope except as survey positioning.

\begin{table}[t!]
  \centering
  \caption{\harsh{Taxonomy of thirteen works. Labeled blocks denote attack family; class denotes access (P$=$physical, S$=$software co-tenant).}}
  \label{tab:taxonomy}
  \fontsize{7.5}{8.1}\selectfont
  \setlength{\tabcolsep}{3.2pt}
  \renewcommand{\arraystretch}{0.94}
  \color{black}\begin{tabular}{@{}lllcc@{}}
    \toprule
    \textbf{Attack} & \textbf{Channel} & \textbf{Platform} & \textbf{Class} & \textbf{Unpriv} \\
    \midrule
    \multicolumn{5}{l}{\textbf{A. Thermal perception integrity}} \\
    Heat is On~\cite{bhupathiraju2026} & IR pipeline & thermal cams & P & -- \\
    Zhu bulbs~\cite{zhu2021} & IR scene & IR detectors & P & -- \\
    HOTCOLD~\cite{wei2023} & IR wearables & IR detectors & P & -- \\
    \midrule
    \multicolumn{5}{l}{\textbf{B. Thermal-telemetry leakage}} \\
    Hot Pixels~\cite{taneja2023} & temp/pwr/freq & Arm SoC/GPU & S & $\bullet$ \\
    ThermalBleed~\cite{kim2022} & coretemp & Intel x86 & S & $\bullet$ \\
    Hertzbleed~\cite{wang2022} & DVFS timing & x86 & S & $\bullet$ \\
    ThermalScope~\cite{zhang2024} & TEI timing & x86 & S & $\bullet$ \\
    Gu et al.~\cite{gu2016} & 3D thermal & 3D IC & S & -- \\
    Knechtel et al.~\cite{knechtel2017} & 3D thermal & 3D IC & S & -- \\
    Masti et al.~\cite{masti2015thermal} & thermal CC & multicore & S & $\bullet$ \\
    \midrule
    \multicolumn{5}{l}{\textbf{C. Model-information extraction}} \\
    PLATYPUS~\cite{murdock2021} & RAPL & x86 & S & \halfcirc \\
    DeepTheft~\cite{gao2024} & RAPL/freq & x86 & S & \halfcirc \\
    BarraCUDA~\cite{horvath2025} & EM & Jetson & P & -- \\
    \bottomrule
  \end{tabular}\\[0.3em]
  {\footnotesize
  $\bullet$: fully unprivileged;
  \halfcirc: partial;
  $\circ$: no;
  --: N/A for class.}
\end{table}


\begin{table}[t!]
  \centering
  \caption{\harsh{Comparison axes for thirteen surveyed works.
  Deploy: H/M/L/U$=$high/medium/low/unknown.
  Rel.: D$=$direct, An$=$analogical, P$=$plausible.}}
  \label{tab:compare-axes}

  \fontsize{7.5}{8.1}\selectfont
  \setlength{\tabcolsep}{3.2pt}
  \renewcommand{\arraystretch}{0.94}

  {\color{black}
  \begin{tabular}{@{}llcccc@{}}
    \toprule
    \textbf{Attack} &
    \textbf{Timescale} &
    \textbf{Co-ten.} &
    \textbf{Deploy} &
    \textbf{Rel.} &
    \textbf{Goal} \\
    \midrule

    Heat is On~\cite{bhupathiraju2026}
      & frame--seconds & $\circ$ & H & D & perception integrity \\

    Zhu bulbs~\cite{zhu2021}
      & scene dwell & $\circ$ & H & D & detector evasion \\

    HOTCOLD~\cite{wei2023}
      & scene dwell & $\circ$ & H & D & detector evasion \\

    Hot Pixels~\cite{taneja2023}
      & seconds & $\bullet$ & M & An & fingerprinting \\

    ThermalBleed~\cite{kim2022}
      & minutes & $\bullet$ & L & An & KASLR leak \\

    Hertzbleed~\cite{wang2022}
      & hours (crypto) & \halfcirc & L & An & key recovery \\

    PLATYPUS~\cite{murdock2021}
      & seconds--hours & $\bullet$ & M & An & power SCA \\

    ThermalScope~\cite{zhang2024}
      & hours / 0.1\,b/s & $\bullet$ & M & An & FP / covert \\

    Gu et al.~\cite{gu2016}
      & design-time & \halfcirc & U & P & packaging leak \\

    Knechtel et al.~\cite{knechtel2017}
      & design-time & \halfcirc & U & P & packaging leak \\

    DeepTheft~\cite{gao2024}
      & offline traces & $\bullet$ & M & An & arch.\ extract \\

    BarraCUDA~\cite{horvath2025}
      & physical probe & $\circ$ & M & An & weight extract \\

    Masti et al.~\cite{masti2015thermal}
      & seconds--minutes & $\bullet$ & L & An & covert channel \\

    \bottomrule
  \end{tabular}

  \vspace{0.3em}

  {\footnotesize
  Co-ten.: $\bullet$ requires shared silicon;
  $\circ$ no;
  \halfcirc\ partial.}
  }
\end{table}

\rev{Despite different access models, the surveyed attacks exploit interfaces built for night perception or thermal reliability: scene inputs affect integrity, while workload observations affect confidentiality.}
\harsh{\rev{The joint comparison identifies which boundary a defense protects: sensor hardening leaves compute observations exposed, and telemetry controls do not authenticate IR inputs. Interactions between perception activity and observable compute load remain hypotheses requiring end-to-end validation.}}
Locking RAPL often leaves frequency or GPU telemetry available
\cite{gao2024,taneja2023,wang2022}, while physical IR attacks bypass compute
entirely~\cite{bhupathiraju2026,zhu2021,wei2023}.

\subsubsection{Key Observation 1.}
Heat was instrumented for night perception or package reliability, not confidentiality.
Physical IR spoofing and unprivileged silicon telemetry therefore form one residual-risk frame for AV designers, yet they do not share defenses, metrics, or evaluation methods.
Closing one access path does not close the other.

\subsection{Physical Thermal Perception}
\label{sec:phys-ir}
Thermal IR sensing remains useful when RGB contrast collapses under night, fog, or glare. The camera pipeline and the external heat field are therefore security-relevant for AV perception.

\emph{Heat is On}~\cite{bhupathiraju2026} analyzes equalization, calibration, and lensing stages in commercial thermal cameras used in autonomous systems (FLIR Boson, InfiRay T2S, FPV XK-C130). Unlike digital adversarial patches on RGB images, the adversary manipulates the radiometric field such that \emph{camera-internal} processing amplifies or suppresses apparent temperatures, including timed artifacts. Equalization and non-uniformity correction, which are essential for usable IR imagery, can amplify localized heat sources into detector-scale blobs. The original evaluation reports mean average precision drops of roughly 50\% for pedestrian detection and 45\% for RGB--thermal fusion. Real-world tests report up to 100\% pedestrian misdetection and 91\% false-obstacle success, including planning-relevant false obstacles that can trigger emergency braking.

Scene-injection work complements pipeline attacks. Low-cost (\$5) bulb boards reduce YOLOv3 infrared pedestrian average precision by 34.5\% in recorded video~\cite{zhu2021}, and wearable HOTCOLD hot/cold patterns alter thermal silhouettes~\cite{wei2023}. Both operate black-box and transfer across detector architectures in the original evaluations. Relative to pipeline attacks, physical patches require less knowledge of camera internals but more control of the scene. Relative to LiDAR spoofing~\cite{cao2019adversarial,shen2021fusion}, they target a modality that AV stacks often trust precisely when RGB fails. Bulb boards favor static roadside placement, whereas wearable HOTCOLD patterns move with a person and can track the ego vehicle. Naive fusion that trusts thermal detections when RGB is dark can allow an IR-only spoof to dominate the fused decision~\cite{bhupathiraju2026}. Related RGB physical attacks~\cite{eykholt2018robust} likewise motivate explicit cross-modal inconsistency checks. \rev{Pipeline attacks motivate camera-processing controls; scene attacks require robustness to plausible thermal silhouettes. Corroboration must account for sensor reliability when RGB fails.} Deployability for AV night sensing is high, yet these attacks are not silicon side channels: defenses and evaluations do not transfer to the software telemetry path.

\subsubsection{Key Observation 2.}
When RGB contrast collapses, AV stacks often elevate thermal detections.
Naive fusion that trusts thermal-only evidence under night or glare allows an IR spoof, including thermodynamically consistent heaters or wearables, to dominate fused decisions and trigger unsafe planner actions~\cite{bhupathiraju2026,zhu2021,wei2023}.

\subsection{Unprivileged Silicon Telemetry}
\label{sec:silicon-telemetry}

\noindent{\textbf{Temperature, power, and frequency.}}
On the silicon path, \emph{Hot Pixels}~\cite{taneja2023} shows that Apple M1, Samsung Exynos, and Qualcomm Snapdragon SoCs expose temperature, power, and clock frequency through unprivileged interfaces, with browser-reachable pixel stealing, history sniffing, and website fingerprinting. Many fingerprint tasks succeed on timescales of seconds, which is short relative to typical OTA update windows. Two observations matter for AVs. First, Snapdragon-class devices appear in automotive and mobile SoC families, indicating that the privilege model is not confined to desktop x86. Second, DVFS is a required design property for thermally constrained accelerators; disabling frequency scaling is not a cost-free patch. The open deployability question is whether an OTA or infotainment co-tenant can fingerprint a perception workload on a shared package under OEM policy. 

\emph{ThermalBleed}~\cite{kim2022} uses Linux \texttt{hwmon} core temperatures to distinguish cache hits from page-table walks and reports KASLR breaks with 76--100\% attack success rate in roughly 6--9\,minutes, depending on configuration. Coarse operating-system sampling is still sufficient to leak memory-access patterns on shared silicon, which is relevant wherever multiple user-space partitions share a package. The attack does not require performance-monitoring units, which matches the unprivileged co-tenant assumptions in Section~\ref{sec:threat-model}. \emph{Hertzbleed}~\cite{wang2022} converts DVFS-induced frequency variation into remote timing, including SIKE key extraction over tens of hours in reported settings. Hertzbleed is included for the shared power--heat--frequency coupling under sustained load, not for remote cryptanalytic relevance. In an AV context, the more immediate concern is local co-tenant observation of frequency under perception load.

\noindent{\textbf{Interrupts and packaging.}}
\emph{ThermalScope}~\cite{zhang2024} times TEIs raised when CPU package temperature crosses manufacturer-determined thresholds. The original evaluation reports covert channels at 0.1\,b/s, DNN architecture fingerprinting above 90\% accuracy, and KASLR recovery in approximately 8.2\,hours. Relative to continuous Hot Pixels-style sampling, TEIs form a thresholded timing channel on the same thermal physics. TEIs cannot be removed by locking a sysfs node without disabling thermal management. Cross-die thermal leakage in 3D integrated circuits, with floorplanning mitigations that reduce Side-channel Vulnerability Factor, is studied by Gu et al.~\cite{gu2016} and Knechtel and Sinanoglu~\cite{knechtel2017}. As AV compute densifies into chiplets, a temperature sensor on a memory die may observe compute-die activity despite logical isolation. That threat remains design-time and is unvalidated on automotive packages.

\vspace{-1.5ex}  
\subsection{Power-Side Model Extraction}
\vspace{-0.8ex}  

\label{sec:model-extract}
Workload fingerprinting identifies which task is running. Model extraction recovers network structure or, in stronger attacks, weights. For AV suppliers, the latter threatens proprietary perception models that may then be subjected to white-box adversarial follow-on.

\emph{DeepTheft}~\cite{gao2024} maps coarse $\sim$1\,kHz RAPL power traces to architecture and hyperparameter recovery, reporting 99.75\% LDA for network-structure recovery and 99.60\% weighted F1 for hyperparameters in the original paper. The same learning framework can instead be trained on CPU frequency traces, which remain accessible on many platforms where RAPL is restricted. Earlier, \emph{PLATYPUS}~\cite{murdock2021} established the broader software RAPL surface, including AES key-byte recovery, fast KASLR breaks, and covert channels near 18.7\,b/s. That work explains subsequent vendor restrictions on \texttt{energy\_uj} and motivates the survey's emphasis on frequency and GPU fallbacks. Classic power-analysis surveys~\cite{lou2021survey,spreitzer2018survey} place both works in an energy-management attack lineage that also includes DVFS fault injection~\cite{tang2017clkscrew,qiu2019voltjockey}.

\emph{BarraCUDA}~\cite{horvath2025} recovers DNN weights via electromagnetic probes on Jetson Nano and Orin. The result strengthens the physical attacker class for model theft on edge GPUs with automotive-adjacent form factors, and it is complementary to software-path architecture recovery. Locking operating-system telemetry does not end model confidentiality risk when physical access to the module is available. For AV supply chains, software architecture shortlisting followed by physical weight extraction is more severe than either path alone: telemetry may reduce the hypothesis space, after which a probe attack targets a smaller set of candidate networks.

\subsubsection{Key Observation 3.}
Restricting a single energy or power interface is necessary but incomplete. Frequency and accelerator telemetry often remain readable and continue to support fingerprinting and architecture shortlisting~\cite{gao2024,taneja2023,wang2022}.
The exposure is a residual telemetry \emph{class}, not a single temperature node.

\section{Defenses and Mitigations}

Closing either access path leaves residual risk on the other. Table~\ref{tab:defense-matrix} maps sensor, telemetry, DVFS, packaging, and model controls to residual risk under the two-class threat model of Section~\ref{sec:threat-model}.
\vspace{-1.5ex}  

\begin{table}[t!]
  \centering
  \caption{Defense coverage vs.\ physical IR and software co-tenant threats.
  Residual-risk phrases are unpacked in this section.
  SOTIF$=$Safety of the Intended Functionality (ISO~21448);
  TEI$=$thermal event interrupt.}
  \label{tab:defense-matrix}
  \fontsize{8.0}{9}\selectfont
  \setlength{\tabcolsep}{3.2pt}
  \renewcommand{\arraystretch}{1.05}
  \begin{tabular}{@{}lccp{3.6cm}@{}}
    \toprule
    \textbf{Defense} & \textbf{Phys} & \textbf{Soft} & \textbf{Residual risk} \\
    \midrule
    Threat-aware IR DSP / fusion~\cite{bhupathiraju2026} & \halfcirc & $\circ$ & Physically consistent heaters still spoof \\
    Temp.\ noise / quantize (observer) & $\circ$ & \halfcirc & Frequency and GPU paths remain \\
    Drop/coarsen GPU power (observer) & $\circ$ & \halfcirc & Temperature and frequency remain \\
    Rate-limit telemetry & $\circ$ & \halfcirc & Long observation windows still leak \\
    Capability-based telemetry policy & $\circ$ & $\bullet$ & TEIs and governors must stay live \\
    DVFS isolate / cap (proposed) & $\circ$ & $\bullet$ & Throughput / SOTIF margin cost \\
    RAPL lockdown & $\circ$ & \halfcirc & Freq/GPU alternatives~\cite{gao2024} \\
    3D floorplanning~\cite{gu2016,knechtel2017} & $\circ$ & $\bullet$ & Applies to new silicon only \\
    Model obfuscation & $\circ$ & \halfcirc & Workload fingerprinting remains \\
    \bottomrule
  \end{tabular}\\[0.3em]
  {\footnotesize
  $\bullet$: strong;
  \halfcirc: partial;
  $\circ$: weak/N/A.}
\end{table}

\vspace{-0.8ex}  
\subsection{Sensor Defenses}
Threat-aware IR DSP~\cite{bhupathiraju2026} and cross-modal consistency checks (thermal versus RGB/LiDAR) reduce opportunistic spoofs that rely on camera-internal equalization amplifying small heat sources. Related perception attacks~\cite{eykholt2018robust,cao2019adversarial,shen2021fusion}
motivate adversarial fusion rather than accuracy-only fusion:
corroboration should be weighted by sensor reliability; absent RGB
support at night is not itself evidence of an IR spoof.

\rev{Operationally, planners should assess thermal-only evidence using reliable cross-modal disagreement, temporal consistency, and detection uncertainty. These checks require planner-level validation and remain incomplete against full scene control.} Large heaters, HOTCOLD wearables~\cite{wei2023}, and bulb boards~\cite{zhu2021} can present thermodynamically consistent IR silhouettes that pass naive physics checks and still fool detectors. Certified robustness offers stronger guarantees, but the inference cost is hard to reconcile with real-time AV budgets. Sensor defenses address the physical path only and do not fully close it; they leave co-tenant telemetry untouched.

\subsection{Telemetry Controls}
Locking thermal sysfs is necessary but insufficient. ThermalBleed and ThermalScope use interfaces tied to management itself~\cite{kim2022,zhang2024}, and RAPL lockdown leaves frequency and GPU power paths~\cite{murdock2021,gao2024,taneja2023}. Observer-side temperature noising or quantization therefore leaves residual frequency and GPU channels, while dropping GPU power leaves temperature and frequency informative (Table~\ref{tab:defense-matrix}).

Rate limiting alone leaves long observation windows informative. Capability-based telemetry policy can restrict Quality Management (QM) and unprivileged readers to coarsened or delayed views while ASIL governors retain full-resolution thermal state; TEIs and DVFS must remain live for package protection. DVFS-domain isolation or caps for perception can reduce observable coupling; however, they trade throughput and SOTIF thermal margin. Model obfuscation may hinder architecture extraction while leaving workload fingerprinting intact. Floorplanning mitigations apply only to new silicon~\cite{gu2016,knechtel2017}.

\subsubsection{Key Observation 4.}
No single defense layer closes both access paths without residual risk.
Sensor controls leave co-tenant telemetry open; restricting one software telemetry channel leaves others informative; thermal interrupts and governors cannot be removed without compromising package thermal safety~\cite{iso21448,zhang2024}.
\section{Evaluation}
\label{sec:evaluation}
\begin{figure}[t!]
  \centering
  \includegraphics[width=\linewidth,trim={70 81 71 80},clip]{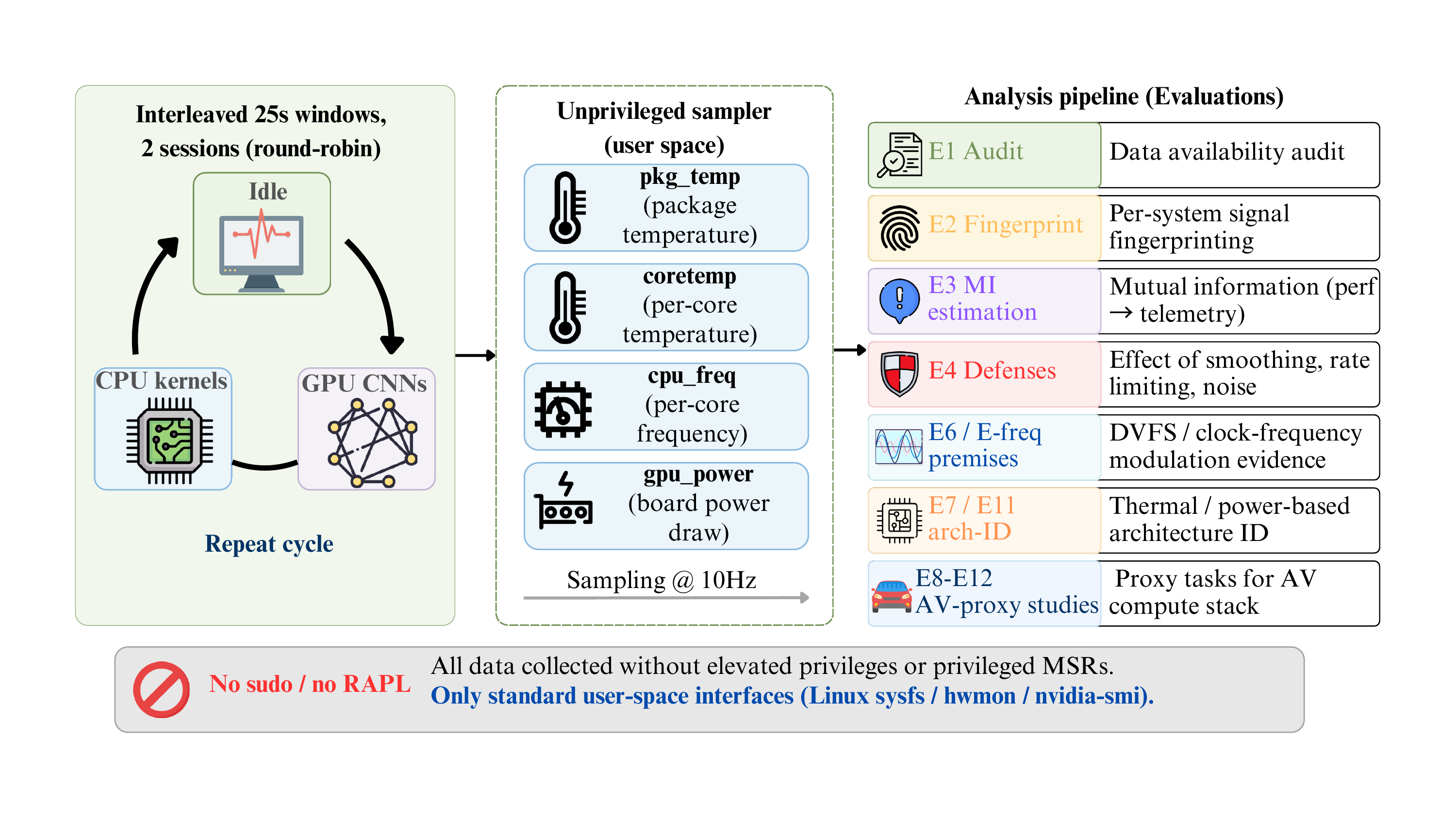}
  \caption{Experimental pipeline for the unprivileged commodity proxy:
  telemetry collection, feature aggregation, fingerprinting, ablation, observer sensitivity, and stress tests.}
  \label{fig:eval-pipeline}
\end{figure}

\rev{This section reports original measurements that test the software-path implications of the survey.} Physical thermal IR attacks already have published empirical results that we cite directly~\cite{bhupathiraju2026,zhu2021,wei2023}, and we do not reproduce physical-path experiments. Public unprivileged telemetry traces for automotive-grade SoCs are unavailable, and the co-tenant results surveyed in Sections~\ref{sec:silicon-telemetry} and~\ref{sec:model-extract} originate from non-automotive platforms. Production DRIVE/Orin/Ride privilege policies are OEM-specific. We therefore treat the following campaign as a \emph{proxy evaluation}: a commodity host under a fully unprivileged policy (no \texttt{sudo}, no RAPL collection) that tests whether residual thermal, DVFS, and GPU telemetry separates AV-analogue workloads when RAPL is locked. \rev{The experiments examine accessibility and redundancy (E1--E3), observer sensitivity (E4), and persistence across sessions, contention, and a null control (E8--E10). Premise tests (E6) and architecture shortlisting (E7/E11) provide narrower links to prior attacks. This mechanism baseline does not estimate in-vehicle fingerprinting rates or validate automotive defenses.} Figure~\ref{fig:eval-pipeline} outlines the experimental pipeline. Table~\ref{tab:experiments} summarizes each experiment; Table~\ref{tab:software-lit} maps software literature claims to the closest proxy outcome.

\begin{table}[t!]
  \centering
  \caption{Proxy experiments (E1--E12) on Xeon\,+\,Blackwell under a fully unprivileged policy (RAPL collection disabled).
  Role: Acc$=$accessibility, Sep$=$separability, Red$=$redundancy, Def$=$defense sensitivity,
  Prem$=$premise check, Ana$=$literature analogue, Ctrl$=$control/null, Lat$=$latency.}
  \label{tab:experiments}
  \fontsize{8}{8}\selectfont
  \setlength{\tabcolsep}{3.5pt}
  \renewcommand{\arraystretch}{1.15}
  \begin{tabular}{@{}llp{3.3cm}p{3.5cm}@{}}
    \toprule
    \textbf{ID} & \textbf{Role} & \textbf{Objective} & \textbf{Primary outcome} \\
    \midrule
    E1 & Acc & Telemetry access audit & Temp/freq/GPU open; RAPL denied \\
    E2 & Sep & 8-class fused fingerprinting & Acc.\ 95.8\%; F1 $0.958$ $[0.92,0.98]$ \\
    E3 & Red & Cross-channel MI & High MI among temp/GPU pairs \\
    E4a & Def & Temp.\ noise / quantize & F1 $\geq 0.948$; CI overlaps baseline \\
    E4b & Def & Drop GPU power (+combined) & F1 $0.842$ (combined $0.822$) \\
    E5 & Ana & Offline RAPL DeepTheft & Skipped (public traces unavailable) \\
    E6a & Prem & Cache vs.\ DRAM $\Delta T$ & $\Delta T{\approx}1.61$\,°C; RF $0.625$ \\
    E6b & Prem & Data-dependent $|\Delta f|$ & $|\Delta f|{\approx}6.0$\,MHz; RF $0.75$ \\
    E7 & Ana & 4-class GPU arch.\ ID & F1 $0.908$ $[0.78,1.0]$; $n{=}32$ \\
    E8 & Sep & Cross-session transfer & Mean transfer F1 $0.941$ \\
    E9 & Sep & CPU/GPU co-tenant stress (4 GPU) & GPU F1 $\rightarrow 0.582$ \\
    E10 & Ctrl & Label-shuffle null & F1 $\approx 0.13$ (chance $0.125$) \\
    E11 & Ana & Top-$k$ arch.\ shortlist & Top-2 accuracy $1.0$ \\
    E12 & Lat & Victim latency under E4 edits & Zero measured overhead \\
    \bottomrule
  \end{tabular}
\end{table}

\subsection{Setup and Methods}

\textbf{Platform:} Intel Xeon w7-3465X with an NVIDIA RTX PRO 6000 Blackwell GPU. Package temperature, mean core temperature, CPU-frequency mean, across-core CPU-frequency standard deviation, and GPU temperature, power, and SM clocks are unprivileged-readable.

\noindent\textbf{Workloads:} Eight AV-analogue victims: CPU idle, dense matrix multiplication, FFT, and AES-style hashing; GPU ResNet-18/50, MobileNetV2, and EfficientNet-B0. These approximate perception compute intensity; they are not production AV detectors or middleware graphs. AES hashing is a crypto-analogue kernel, not an AV perception task.

\noindent\textbf{Collection:} We collect interleaved 25\,s windows at 10\,Hz with 12 repeats, at least 30\,s cooldown, and two sessions, yielding $n{=}192$ windows for the main fingerprint task. Features are per-window aggregates (mean, standard deviation, minimum, maximum, slope) over seven telemetry series (35 dimensions). The classifier is a random forest with stratified 5-fold cross-validation and bootstrap 95\% confidence intervals. The classifier tests separability of the residual channel class, not attack optimality against adaptive defenses.

\begin{table}[t!]
  \centering
  \caption{Software co-tenant literature versus closest proxy outcome.
  Tags [C]/[P]/[A]/[L]: characterization / premise / analogue / literature-only.}
  \label{tab:software-lit}
  \fontsize{8.0}{9}\selectfont
  \setlength{\tabcolsep}{3.2pt}
  \renewcommand{\arraystretch}{1.05}
  \begin{tabular}{@{}lp{3.1cm}p{3.9cm}@{}}
    \toprule
    \textbf{Attack} & \textbf{Original claim} & \textbf{Closest proxy outcome} \\
    \midrule
    Hot Pixels~\cite{taneja2023} & Browser FP ASR on Arm SoCs & E2 multi-channel separability\,[C] \\
    ThermalBleed~\cite{kim2022} & KASLR 76--100\% & E6a $\Delta T{\approx}1.61$\,°C; RF $0.625$\,[P] \\
    Hertzbleed~\cite{wang2022} & SIKE key; 36--89\,h & E6b $|\Delta f|{\approx}6.0$\,MHz; RF $0.75$\,[P] \\
    PLATYPUS~\cite{murdock2021} & AES/KASLR/covert via RAPL & RAPL denied\,[L] \\
    ThermalScope~\cite{zhang2024} & DNN FP $>$90\% via TEI & No TEI harness\,[L] \\
    DeepTheft-a~\cite{gao2024} & RAPL struct.\ LDA 99.75\% & E5 skipped (no public traces)\,[L] \\
    DeepTheft-b~\cite{gao2024} & Arch.\ shortlisting goal & E7 GPU F1 $0.908$\,[A]; E11 top-2 $1.0$ \\
    \bottomrule
  \end{tabular}\\[0.3em]
  {\footnotesize TEI$=$thermal event interrupt.}
\end{table}

\subsection{Fingerprinting and Ablation}

\noindent\textbf{E1.} Accessibility audit under a fully unprivileged policy: RAPL energy counters in microjoules (\texttt{energy\_uj}) are present but denied, while temperature, frequency, and GPU telemetry remain open.

\begin{figure}[t!]
  \centering
  \includegraphics[width=0.65\linewidth]{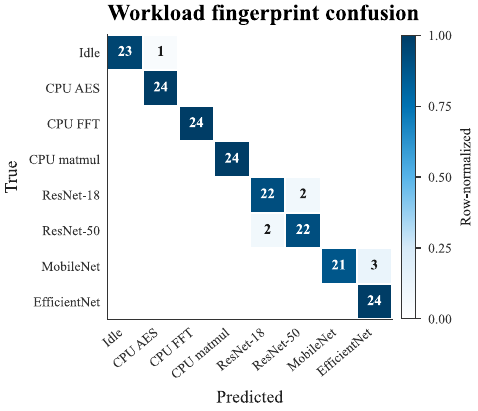}
  \caption{E2 workload fingerprint confusion matrix (row-normalized) for eight AV-analogue classes. } 
  \label{fig:confusion}
\end{figure}

\begin{figure}[t!]
  \centering
  \includegraphics[width=0.65\linewidth]{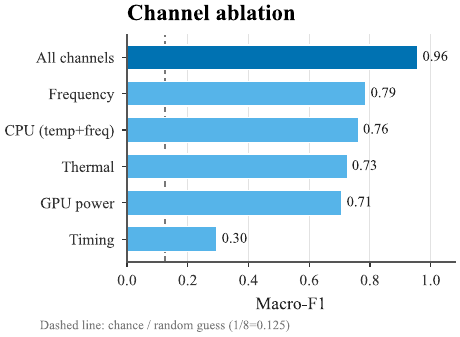}
  \caption{E2 channel ablation: macro-F1 by telemetry feature group
  (All $0.958$; Freq.\ $0.787$; CPU $0.762$; Thermal $0.726$; GPU power $0.709$; Timing $0.297$).
  The dashed line marks chance under uniform guessing over eight classes ($1/8=0.125$).}
  \label{fig:ablation}
\end{figure}

\noindent\textbf{E2.} Stratified cross-validation accuracy is 95.8\% with macro-F1 $0.958$ $[0.92,0.98]$. Figure~\ref{fig:confusion} shows the corresponding confusion matrix. Fused temperature, frequency, and GPU features dominate (Figure~\ref{fig:ablation}): frequency-only $0.787$, thermal-only $0.726$, GPU-power-only $0.709$, and timing-only $0.297$ (weakest channel; chance $0.125$). \harsh{On the commodity proxy, fused telemetry features yield higher workload-fingerprinting macro-F1 than any individual feature group.}

\noindent\textbf{E3.} Mutual information among channels is non-trivial (package versus core temperature $1.29$\,bits; GPU power versus SM clock $0.99$\,bits), consistent with redundant leakage paths.

\subsection{Observer Sensitivity}
\begin{figure}[t!]
  \centering
  \includegraphics[width=\linewidth]{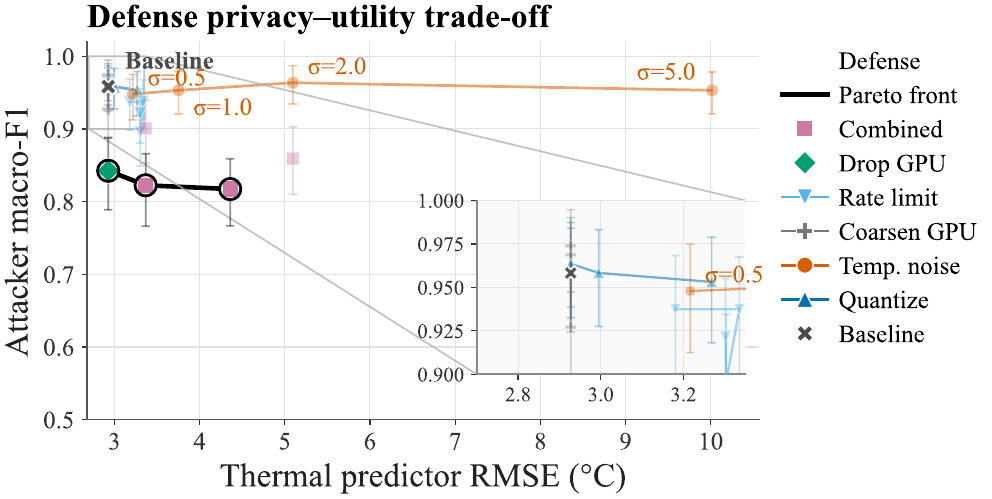}
  \caption{E4 observer-side sensitivity: attacker macro-F1 versus thermal-predictor RMSE.
  Temperature noising alone leaves attack F1 high; dropping GPU power and combined policies move along the Pareto front.}
  \label{fig:pareto}
\end{figure}

\textbf{E4.} Temperature noise, quantization, rate limiting, and GPU-power feature dropping are applied to the attacker feature vector. This is observer-side feature degradation, not an operating-system capability policy. Temperature noising (E4a) raises thermal-predictor RMSE ($2.93\rightarrow 10.02$\,°C at $\sigma{=}5$) while all-channel F1 remains high: confidence intervals overlap the baseline at every noise level, with the lowest point estimate $0.948$ at $\sigma{=}0.5$ (Table~\ref{tab:pareto}). Frequency and GPU power are untouched by temperature noise; the fused classifier therefore continues to separate workloads, matching the defense-matrix residual risk that frequency and GPU paths remain. Dropping GPU power (E4b) reduces F1 to $0.842$; a combined quantize, rate-limit, and drop policy reaches $0.822$ (Table~\ref{tab:pareto} and Figure~\ref{fig:pareto}), with confidence intervals excluding the baseline. Temperature and frequency remain informative after GPU power is removed. No operating point jointly yields low attack F1 and low RMSE.

\begin{table}[t!]
  \centering
  \caption{E4 observer-side sensitivity (merged).
  Acc/F1: attacker fingerprinting; RMSE: temperature predictor.
  $^{\ast}$CI overlaps baseline; $^{\dagger}$CI excludes baseline.}
  \label{tab:pareto}
  \fontsize{8.0}{9}\selectfont
  \setlength{\tabcolsep}{3.5pt}
  \renewcommand{\arraystretch}{1.15}
  \begin{tabular}{@{}lcccc@{}}
    \toprule
    \textbf{Setting} & \textbf{Acc} & \textbf{F1} & \textbf{RMSE (°C)} & \textbf{$\Delta$F1} \\
    \midrule
    Baseline & 0.958 & 0.958 & 2.93 & -- \\
    Gauss.\ $\sigma{=}0.5$ & 0.948 & 0.948 & 3.22 & $-0.01^{\ast}$ \\
    Gauss.\ $\sigma{=}2.0$ & 0.964 & 0.964 & 5.10 & $+0.01^{\ast}$ \\
    Gauss.\ $\sigma{=}5.0$ & 0.953 & 0.953 & 10.02 & $-0.01^{\ast}$ \\
    Drop GPU power & 0.844 & 0.842 & 2.93 & $-0.12^{\dagger}$ \\
    Quantize $2$\,°C $+$ rate $1$\,Hz $+$ drop GPU & 0.823 & 0.822 & 3.37 & $-0.14^{\dagger}$ \\
    \bottomrule
  \end{tabular}
\end{table}

\subsubsection{Key Observation 5.}
On the commodity proxy, temperature-only observer noising leaves attack accuracy high (confidence intervals overlap the baseline), while dropping GPU-power features measurably reduces fingerprinting performance.
No operating point jointly yields low attack accuracy and low thermal-predictor error (Table~\ref{tab:pareto}).

\subsection{Stress Tests and Premises}

\textbf{E9 (co-tenant stress).} 
On the four GPU victims (distinct from the eight-class E2 task), four-class solo fingerprint F1 is $0.958$ $[0.85,1.0]$; under a GPU co-tenant it falls to $0.582$ $[0.34,0.76]$, and a CPU co-tenant yields $0.829$ $[0.61,0.96]$. The wider solo CI reflects the smaller four-class subset, not a revision of the eight-class E2 interval. Mixed-criticality contention weakens the solo-host result. Solo F1 provides a quiet-host baseline; quieter partitions or longer observation windows may still leave residual leakage.

\noindent\textbf{E8 (cross-session transfer).}
Mean transfer F1 is $0.941$ (drop $\approx$0.017 relative to within-session evaluation), indicating that the fingerprint remains stable across the two collection sessions.

\noindent\textbf{E10 (null control).}
Label-shuffle null F1 is $\approx$0.13, matching chance ($0.125$) for eight classes.

\begin{figure}[t!]
  \centering
  \includegraphics[width=0.55\linewidth]{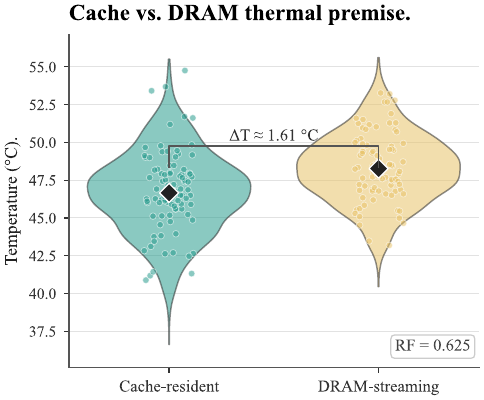}
    \caption{E6a ThermalBleed-style premise: package-temperature distributions for cache-resident versus DRAM-streaming workloads ($\Delta T{\approx}1.61$\,°C; binary RF $0.625$).}
  \label{fig:thermal-distinguish}
\end{figure}

\noindent\textbf{E6a (ThermalBleed premise).}
Cache versus DRAM thermal distinguishability yields $\Delta T \approx 1.61$\,°C and RF accuracy $0.625$ on a binary task (Figure~\ref{fig:thermal-distinguish}).

\noindent\textbf{E6b (Hertzbleed premise).}
High versus low Hamming-weight frequency yields $|\Delta f| \approx 6.0$\,MHz and RF accuracy $0.75$.

\noindent\textbf{E5, E7, and E11 (DeepTheft mapping).}
Offline RAPL DeepTheft (E5) was skipped because public RAPL training traces matching the original setting were unavailable. As a narrower analogue of the DeepTheft architecture-identification goal, E7 trains a four-class GPU-telemetry classifier on toy networks and reports macro-F1 $0.908$ $[0.78,1.0]$ ($n{=}32$). E11 then shows that top-2 shortlisting reaches accuracy $1.0$ ($2\times$ reduction versus a uniform four-class prior).

\noindent\textbf{E12 (victim latency).}
Observer-side feature edits add no measured overhead to victim inference; E4 measures their effect on thermal-prediction error, not on governor accuracy or thermal safety.

\section{Discussion and Future Research Direction}
\label{sec:discussion}
The evaluation supports the software-path half of the taxonomy under a commodity privilege model. This section interprets those results for transfer, states limitations explicitly, derives engineering obligations, and enumerates open problems with concrete research directions.

\subsection{Limitations and Transfer}
Measurements use a single commodity host and cooling regime, not Jetson, DRIVE, or Ride platforms. Physical IR, TEI, KASLR, SIKE, and live RAPL DeepTheft reproductions are out of scope. E6a and E6b are weak premises and must not be read as confirming the full attacks in the cited papers. Thermal drift and boost-clock history remain residual confounds despite interleaving, dual sessions, and null controls. Feature vectors are window aggregates; sample-level cryptanalytic distinguishers are not claimed.

Under GPU co-tenancy on the four GPU victims, fingerprint F1 falls from four-class solo macro-F1 $0.958$ to $0.582$ (E9), which constrains rather than strengthens quiet co-tenant claims. The software-path mechanism nonetheless remains plausible wherever (i)~computation yields workload-dependent power and heat, (ii)~governors couple that dissipation to observable frequency or interrupts, and (iii)~weakly privileged software can read those signals, as Hot Pixels reports on Arm and Snapdragon-class devices~\cite{taneja2023}. Observer-side temperature noising raises thermal-predictor RMSE while leaving attack F1 high; dropping GPU-power features lowers F1 to $0.842$, and a combined policy reaches $0.822$ (Table~\ref{tab:pareto}). E7/E11 show architecture shortlisting from residual GPU telemetry after RAPL lockdown, consistent with an inferred DeepTheft-style escalation path~\cite{gao2024}. Absolute fingerprinting rates on automotive SoCs remain unvalidated; the proxy is a mechanism baseline, not an on-vehicle risk estimate.

\subsection{Engineering Implications}

\rev{The surveyed attacks motivate controls beyond locking a single sysfs node. Channel substitution and E4's residual fingerprinting favor telemetry inventories, while E9 makes contention part of the risk assessment. Physical-path recommendations rely on cited IR studies. These implications concern interface design and residual exposure, not transferring the proxy's absolute F1 to AVs.}
\noindent\paragraph{\textbf{Physical IR integrity.}}
\rev{The surveyed pipeline and scene attacks motivate reliability-aware RGB--thermal fusion under RGB collapse~\cite{bhupathiraju2026}. Plausible IR spoofs can dominate decisions when thermal evidence is trusted without considering attack uncertainty. Engineering tests should evaluate cross-modal and temporal consistency, threat-aware DSP, and planner responses to heaters, bulb boards, and HOTCOLD patterns~\cite{zhu2021,wei2023}. Validation should measure spoof resistance and missed genuine hazards when corroborating sensors are unreliable.} Sensor defenses address the physical path only and leave co-tenant telemetry open.

\noindent\paragraph{\textbf{Telemetry hygiene.}}
Restricting RAPL energy counters is common and insufficient~\cite{gao2024,murdock2021}. A per-SoC \emph{channel inventory} should record, for each interface, the principal (governor, ASIL partition, QM partition, unprivileged process), permitted rate, resolution, and retention. Without that inventory, freedom-from-interference claims cannot assert that QM software cannot observe ASIL thermal state. Operational controls follow from the inventory: capability-based reads, coarsened or delayed views for non-safety partitions, and audit of high-rate sampling from OTA or infotainment processes. Observer-side temperature noising alone does not substitute for these controls.

\noindent\paragraph{\textbf{Dual-use thermal control.}}
    Governors and TEIs are required for package protection and for sustaining perception clocks under thermal load; the same mechanisms are observation points~\cite{zhang2024}. Mitigation must preserve ASIL thermal control accuracy while restricting the view available to QM partitions. Removing TEIs or disabling DVFS to close a side channel is not a viable product posture under SOTIF / ISO~21448 thermal envelopes~\cite{iso21448}. DVFS-domain isolation for perception should be evaluated as a design alternative with explicit latency and headroom budgets. 

\noindent\paragraph{\textbf{Standards mapping.}}
Under ISO~26262~\cite{iso26262}, freedom-from-interference (FFI) analysis should treat unprivileged thermal and power telemetry as a potential QM$\leftrightarrow$ASIL exfiltration path whenever mixed-criticality partitions share a package. Evidence for interference need not be a full cryptographic break: workload fingerprinting that reveals which perception model is running is already a confidentiality concern. SOTIF~\cite{iso21448} informs validation of perception insufficiencies and defense-induced performance limitations; deliberate IR manipulation
is addressed through cybersecurity risk assessment. ISO/SAE~21434~\cite{iso21434} Threat Analysis and Risk Assessment (TARA) can list readable telemetry interfaces as assets, co-tenant sampling as an attack path, and channel-inventory gaps as vulnerabilities. These are recommended process mappings, not claims that current standards text already enumerates the channels.

\subsection{Open Problems}
Priority gaps follow from the taxonomy, the defense residual-risk matrix, and the proxy limitations. For each problem we note a concrete research direction.

\begin{itemize}
  \item \textbf{On-target co-tenant evaluation.}
  End-to-end fingerprinting and architecture shortlisting on Orin, DRIVE, or Ride perception stacks under production privilege models remain unmeasured. A natural next step is to port the E1--E12 harness to automotive BSP images, with OEM-approved telemetry inventories and realistic OTA/infotainment co-tenants as observers.

  \item \textbf{Fusion trust under IR spoofing.}
  Night performance still conflicts with IR spoof resistance~\cite{bhupathiraju2026,zhu2021,wei2023}. Future work should quantify planner-level policies that weight RGB/LiDAR corroboration by sensor reliability, and should evaluate threat-aware DSP against thermodynamically consistent heaters rather than accuracy-only fusion metrics.

  \item \textbf{Chiplet and 3D thermal coupling.}
  Cross-die leakage arguments from Gu et al.\ and Knechtel and Sinanoglu~\cite{gu2016,knechtel2017} are design-time and unvalidated on automotive packages. Packaging studies that instrument memory-die sensors under perception load, together with floorplanning SVF reductions under AV thermal envelopes, would close this gap.

  \item \textbf{Capability-based telemetry and SOTIF budgets.}
  Capability policies that coarsen QM views while preserving ASIL governor fidelity lack open reference implementations. Research should measure the latency and thermal-margin cost of DVFS-domain isolation and delayed/coarsened sysfs views under ISO~21448 envelopes~\cite{iso21448}.

  \item \textbf{Standards-ready TARA artifacts.}
  OEM threat analyses rarely enumerate readable thermal interfaces as assets. Publishing checklist templates that map channel inventories onto ISO~26262 FFI and ISO/SAE~21434 TARA~\cite{iso26262,iso21434}, with residual-risk language tied to TEI/DVFS liveness, would accelerate adoption.

 \item \textbf{Contention as a defensive signal.}
  E9 shows that GPU co-tenancy can erase much of the solo fingerprinting advantage. Systematic study of whether scheduled mixed-criticality noise can be \emph{designed} as a soft mitigant, without harming perception deadlines, complements OEM policies that close telemetry at the source. 
\end{itemize}
\section{Conclusion}

This survey systematized thermal IR perception attacks and unprivileged thermal/DVFS/TEI telemetry across physical and software co-tenant classes. Comparing \harsh{thirteen} works by privilege, timescale, co-tenancy, deployability, and relevance confidence, we showed that the paths are complementary integrity and confidentiality failures of interfaces AV platforms must keep operational for night perception and package thermal safety. Closing one path does not close the other; defenses and metrics do not transfer.
Five key observations synthesize the literature and software-path proxy: heat was instrumented for reliability or night perception, not confidentiality; naive thermal-only fusion under RGB collapse can allow an IR spoof to dominate fused decisions; restricting a single energy interface leaves residual frequency and accelerator telemetry informative; no single defense closes both access paths while thermal governors stay live; and temperature noising alone does not eliminate fingerprinting when alternative channels remain open. These observations motivate joint residual-risk reasoning, even though each path still needs separate mitigations.

The broader impact is a residual-risk frame linking sensor, OS, silicon, and standards concerns that prior surveys leave fragmented. Progress requires on-target co-tenant evaluation, fusion trust under IR spoofing, chiplet thermal coupling, capability-based telemetry under SOTIF budgets, and standards-ready TARA artifacts. Through this survey, we aim to raise awareness of thermal interfaces as a first-class AV security surface and to encourage co-design across perception, energy management, and automotive safety.

%
%
%
\bibliographystyle{splncs04}
\bibliography{ref.bib}
\end{document}